# Decoding PPP Corrections from BDS B2b Signals Using a Software-defined Receiver：an Initial Performance Evaluation

**Xiangchen Lu[1], Liang Chen[1], Nan Shen[1], Lei Wang[1], Zhenhang Jiao[1], and Ruizhi Chen[1]**

[1] State Key Laboratory of Information Engineering in Surveying, Mapping and Remote Sensing, Wuhan University, Wuhan 430079, China

Corresponding author: Liang Chen (e-mail: l.chen@whu.edu.cn).

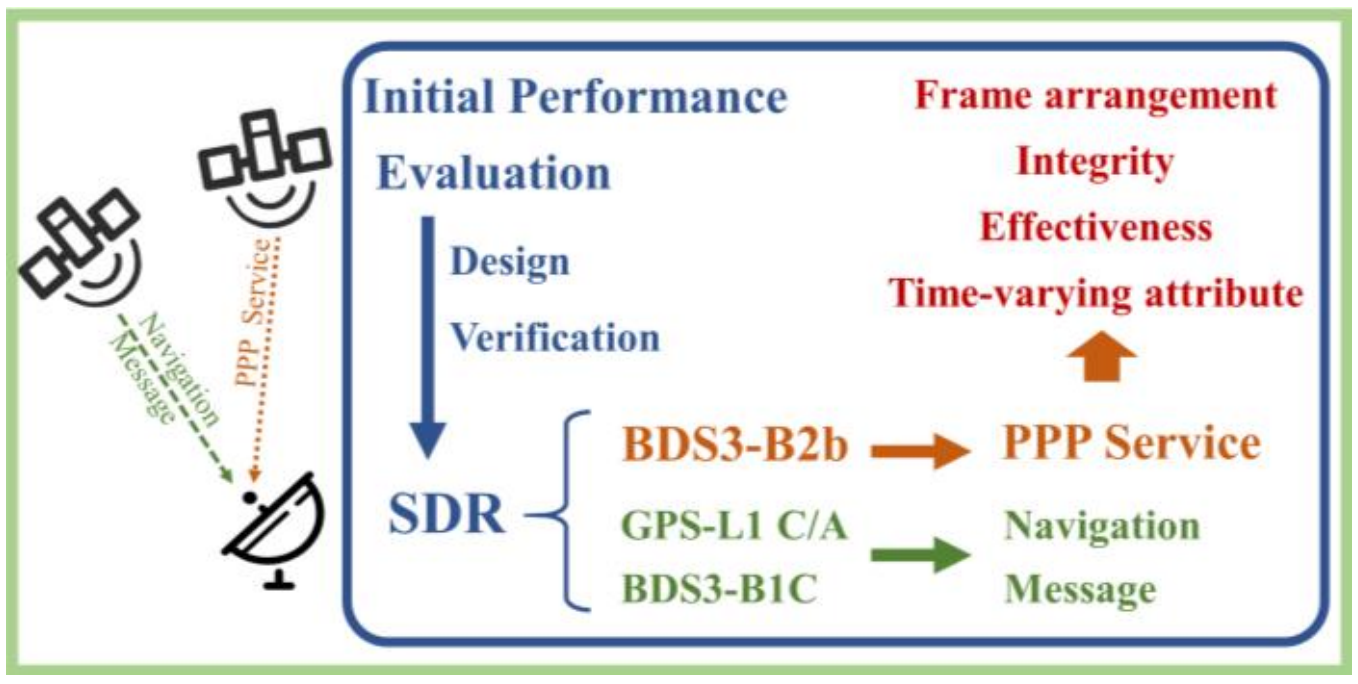

**ABSTRACT** With the rapid development of China's BeiDou Navigation Satellite System (BDS), the application of real-time precise point positioning (RTPPP) based on BDS has become an active research area in the field of Global Navigation Satellite Systems (GNSS). BDS has provided the service of broadcasting RTPPP information. It indicates that BDS has become the second satellite system that provides RTPPP services, following Galileo among the GNSS, but work based on this direction has yet to be explored. Therefore, this paper evaluates the performance of precise point positioning (PPP) service using a software-defined receiver (SDR). An experiment was carried out to verify the feasibility of the SDR. The PPP-B2b signal was processed to obtain PPP service information, including orbit corrections, clock corrections, and differential code bias corrections. The time-varying attributes of these corrections of BDS and GPS are evaluated, and the integrity and stability of the PPP service were analyzed. The results show the PPP-B2b signal can stably provide PPP services for satellites in the Asia-Pacific region, including centimeter to decimeter-level orbit corrections and meter-level clock corrections for BDS satellites. At the same time, PPP services provide decimeter to meter-level orbit correction and meter-level clock correction for GPS satellites. Finally, detection tip for bitstream availability in SDR is proposed. Some content which is not defined in the official document, such as the PPP-B2b frame arrangement, various correction update cycles and the progress of PPP service are discussed.

**INDEX TERMS** Real-time precise point positioning (RTPPP); BeiDou Navigation Satellite System (BDS); Signal processing; Software-defined receiver (SDR).

## I. INTRODUCTION

The development of Chinese BeiDou navigation satellite system (BDS) can be divided into three steps. The first step was to construct the BeiDou Satellite Navigation Demonstration System (BDS-1) [1]. Using an active positioning scheme, the system provided users in China with positioning, timing, wide-area differential and short message communication services. Since 2003, the third BeiDou navigation experiment satellite was launched, further enhancing the performance of the BeiDou Navigation Satellite Demonstration System [2]. The second step was the construction of the BeiDou Satellite Navigation Regional System (BDS-2). In addition to a technical scheme compatible with that of BDS-1, BDS-2 further included a passive positioning scheme, and provided users in the Asia-Pacific region with positioning, velocity measurement, timing, and short message communication services [1]. By the end of 2012, a total of 14 satellites, including five Geostationary Earth Orbit (GEO) satellites, five Inclined GeoSynchronous Orbit (IGSO) satellites, and four Medium Earth Orbit (MEO) satellites, had been launched to complete the space constellation deployment [3]. The third step was to construct the BeiDou Satellite Navigation System (BDS-3) [1]. A total of 19 satellites have been launched to complete a preliminary system for global service [4]. There are plans to complete the deployment of BDS-3 comprehensively with the launching of 30 satellites by 2020 [3]. On June 29, 2020, in orbit testing and network access evaluation of the 55th satellite of BDS was completed, which marks the formal establishment of BDS-3. BDS-3 by inheriting the technical schemes of both active and passive services, can provide global users with positioning, navigation and timing, global short message communication, and international search and rescue services. At the same time, BDS-3 adds regional services for users in China and surrounding areas, including satellite-based augmentation, ground augmentation and precise point positioning (PPP) services, etc [4], [5].

Since the concept of Precise Point Positioning (PPP) was first proposed, it has been a hot research direction in the field of GNSS. However, the precise products that the PPP relies on often have a time delay. In order to meet the demand for real-time precise point positioning (RTPPP), the International GNSS Service (IGS) since 2013 officially has provided real-time precision products (RTS), which are estimated by the analysis centers (ACs) [6]-[8]. The RTS are transmitted around the world using the network transmission of the Maritime Radio Technical Committee (RTCM) under the Internet Protocol (Ntrip) [9]. In addition, several ACs, for example, the German Aerospace Center (DLR), German Research Centre for Geosciences (GFZ), Federal Agency for Cartography and Geodesy (BKG), and Centre National d'Etudes Spatiales (CNES) also provide such corrections, which can be accessed and processed via BKG NTRIP client (BNC) or real-time kinematic library (RTKLIB) software on the user side. With the rapid development of BDS, BDS-based RTPPP has been receiving more attention. Since the CNES became the first organization to provide RTS for BDS satellites in November 2015, organizations such as GFZ and DLR have also begun to release RTS for BDS satellites [10]-[13].

In addition to the RTS provided by IGS, Galileo has been providing High Accuracy Services (HAS) for professional users at a positioning accuracy of 20 cm. HAS corrections will be disseminated through the Galileo E6-B signal on the 1278.75 MHz frequency by MEO satellite. The received HAS corrections can be decoded with a success rate of 90% in an occluded environment [14]-[17]. Japanese QZSS-L6D signals provide a Centimeter Level Augmentation Service (CLAS) for Japanese mainland regions, covering GPS, QZSS, and Galileo systems [18]. In August 2020, BDS released the PPP-B2b document to provide RTPPP service. The joining of this service announced that ordinary users of BDS can obtain RTPPP service by receiving PPP-B2b signals only, without RTCM or BNC.

At present, the receiver for PPP-B2b signal processing is not widely used, and the progress and performance of PPP service provided by the PPP-B2b signal has not been evaluated. To fill this gap, we carried out a preliminary study to explore and analyze the PPP service broadcast in the PPP-B2b signal. The implementation of this work depends on the high flexibility and easy operation of a software-defined receiver (SDR) [19], [20].

The main contributions of this paper are listed below:

1) A complete SDR solution on decoding PPP service information is proposed, and all the steps of SDR, including modules of acquisition, tracking, bitstream and PPP information extraction, are described in detail;
2) A frame arrangement of complete PPP service information is provided, through which the update interval of various PPP corrections can be obtained;
3) The clock and orbit corrections for BDS and GPS are compared and evaluated, and the integrity and stability of the PPP service are analyzed.

The paper is organized as follows: information about the PPP-B2b signal, including signal characteristics, encoding modulation, and navigation message structure, is introduced in Section II. The design scheme and decoding process of the SDR based on the B2b signal are presented in Section III. The experiments and results are presented next, followed by a discussion of the optimization scheme in the decoding process and the characteristics of the PPP service, and the conclusions.

## II. THE STRUCTURE OF PPP-B2b SIGNALS AND PPP MESSAGES

### A. SIGNAL STRUCTURE

The format of the PPP-B2b signal is given in the official BDS documents. As one of the BDS3 signals, the B2b signal is broadcast by all BDS-3 satellites (MEO/ IGSO/ GEO) with a center frequency of 1207.14 MHz and bandwidth of 20.46 MHz. Table I shows the B2b signal structure and broadcast content based on different satellite types [21].

TABLE I

STRUCTURE OF B2b SIGNAL

| Signal | B2b | PPP-B2b |
|---|---|---|
| Content | Position Service | PPP Service |
| Satellite type | MEO/IGSO | GEO |
| Component | I | |
| Carrier frequency (MHz) | 1207.14 | |
| Modulation | BPSK(10) | |
| Symbol rate (sps) | 1000 | |

As shown in Table I, the center frequency of 1207.14MHz signal can be divided into B2b and PPP-B2b signals for BDS-3. Both signals only broadcast the in-phase component I, and their modulation and symbol rates are the same. But MEO and IGSO satellites of the BDS-3 broadcasts the navigation message for position service. The PPP service is provided by the first three GEO satellites on the BDS-3 system. The influence of BPSK on the Gold code, PPP service data and carrier of the PPP-B2b signal is shown in Fig. 1.

As shown in Fig. 1, the PPP-B2b signal $s_{\mathrm{B2b_I}}(t)$ is generated by modulating the PPP service data $D_{\mathrm{B2b_I}}(t)$ and the ranging code $C_{\mathrm{B2b_I}}(t)$. The mathematical expression of $s_{\mathrm{B2b_I}}(t)$ is as follows [21]:

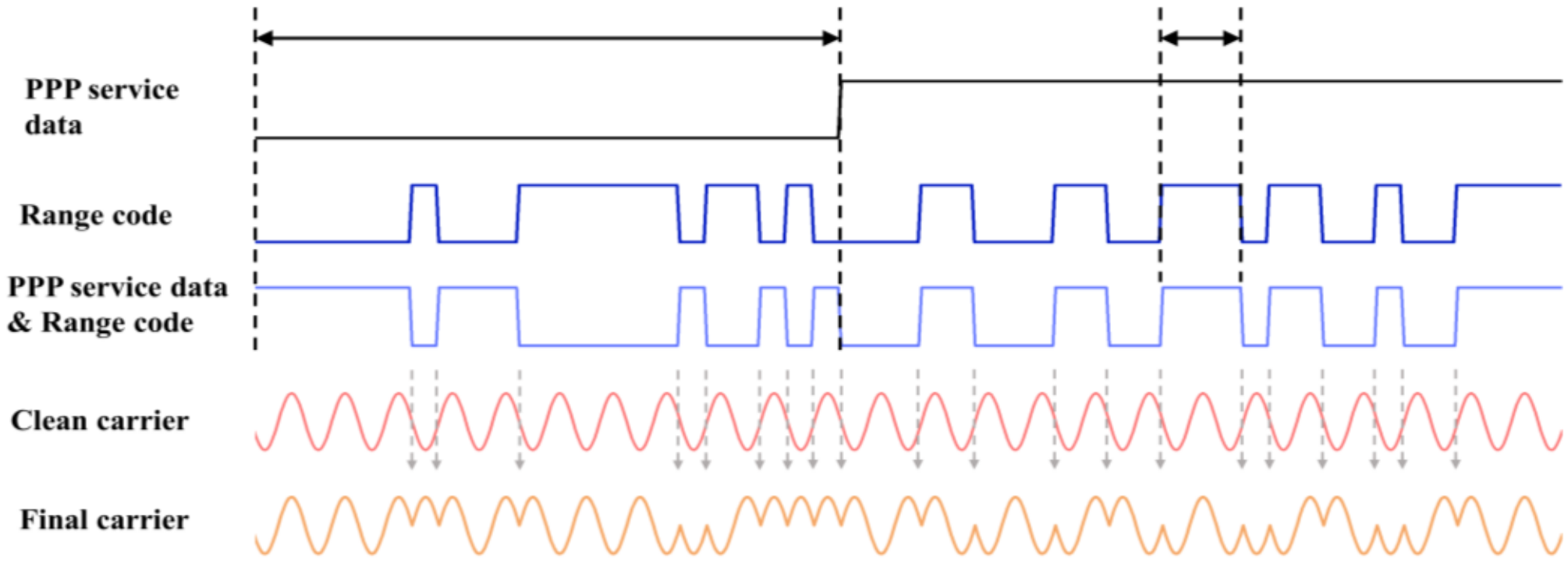

Fig. 1. Influence of BPSK modulation on Gold code, PPP service data, and B2b carrier

$$s_{\mathrm{B2b_I}}(t) = \frac{1}{\sqrt{2}} D_{\mathrm{B2b_I}}(t) \cdot C_{\mathrm{B2b_I}}(t) \tag{1}$$

where $t$ denoted that the signal is continuous in time.

$D_{\mathrm{B2b_I}}(t)$ is as follows [21]:

$$D_{\mathrm{B2b_I}}(t) = \sum_{k=-\infty}^{\infty} d_{\mathrm{B2b_I}}[k] p_{T_{\mathrm{B2b_I}}}(t - kT_{\mathrm{B2b_I}}) \tag{2}$$

where $k$ is the chip number of the corresponding data code; $d_{\mathrm{B2b_I}}$ is the PPP service data code; $T_{\mathrm{B2b_I}}$ is the chip width of the corresponding data code; $p_{T_{\mathrm{B2b_I}}}(t)$ is a rectangle pulse with width of $T_{\mathrm{B2b_I}}$ .

The mathematical expression of range code $C_{\mathrm{B2b_I}}(t)$ is as follows [21]:

$$C_{\mathrm{B2b_I}}(t) = \sum_{n=-\infty}^{\infty} \sum_{k=0}^{N_{\mathrm{B2b_I}}-1} c_{\mathrm{B2b_}I}[k] p_{T_{c_\mathrm{B2b_I}}}\left(t - \left(N_{\mathrm{B2b_I}} n + k\right) T_{c_\mathrm{B2b_I}}\right) \tag{3}$$

where $n$ is the number of the broadcasted ranging code; $N_{\mathrm{B2b_I}}$ is the ranging code length with a value of 10230; $T_{c_\mathrm{B2b_I}} = 1/R_{c_\mathrm{B2b_I}}$ is the PPP-B2b chip period of the ranging code, and $R_{c_\mathrm{B2b_I}} = 10.23$ Mbps is the PPP-B2b_I chipping rate; and $p_{T_{c_\mathrm{B2b_I}}}(t)$ is a rectangle pulse with a duration of $T_{c_\mathrm{B2b_I}}$[21]. Among them, 10 ranging codes are defined for PPP-B2b. Each ranging code $c_{\mathrm{B2b_I}}[k]$ is obtained by expanding the Gold code that is generated by the modulo-2 addition of the shifted output of the two 13-stage linear feedback shift registers [21]. The introduction of this part is to understand the signal structure. The next part is the message structure of PPP information, which is essential for extracting PPP information.

### B. PPP MESSAGE

Each PPP message has a length of 486 bits, wherein the highest 6 bits indicate the message type, the lowest 24 bits are cycle redundancy check (CRC), and the remaining 456 bits are message data, as shown in Fig. 2.

As represented in Fig. 2, each PPP message has a length of 486 bits and is encoded into 972 symbols through 64-ary Low Density Parity Check (64-ary LDPC). These symbols are concatenated together with 16 symbols of the preamble, 6 symbols for the PRN and 6 symbols for the reserved flags to form one PPP-B2b navigation frame with the total length of 1000 symbols. Each PPP message includes 6 bits of message type (Mestype), 456 bits of message data and 24 bits of CRC. The specific contents of message

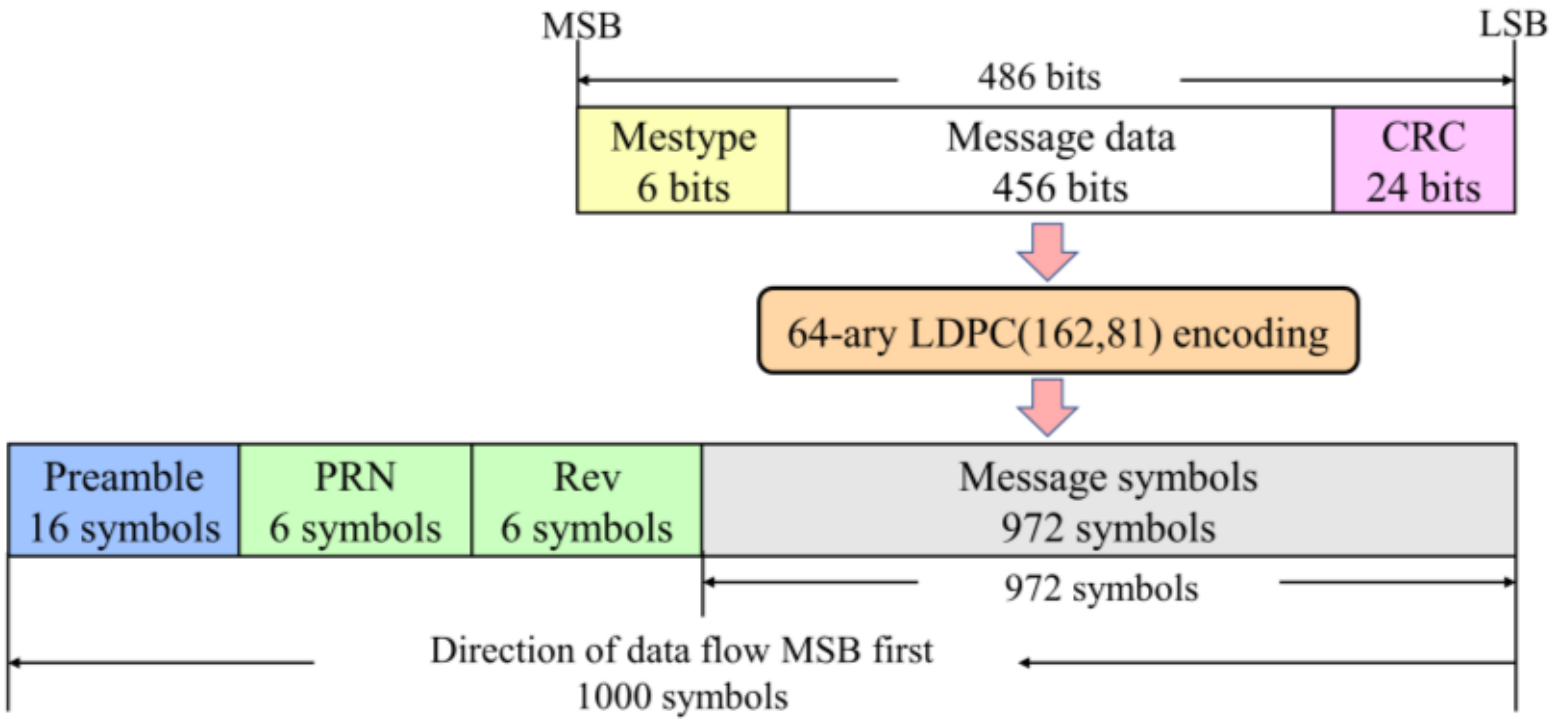

Fig. 2. Frame structure and channel coding of PPP-B2b navigation information

TABLE II

DEFINED MESSAGE TYPES

| **Message types (in decimal)** | **Information content** |
|---|---|
| 1 | Satellite mask |
| 2 | Satellite orbit correction and user range accuracy index |
| 3 | Differential code bias |
| 4 | Satellite clock correction |
| 5 | User range accuracy index |
| 6 | Clock correction and orbit correction – combination 1 |
| 7 | Clock correction and orbit correction – combination 2 |
| 8-62 | Reserved |
| 63 | Null message |

data may depend on the different type of Mestype, which is listed in Table II. Currently, seven categories (Mestype 1-7) have been put into use and the remaining messages (Mestype 8-62) are reserved. The Null message (Mestype 63) are filled in the blank time [21].

## III. BDS SOFTWARE-DEFINED RECEIVER

SDR has a flexible open-architecture, which can be easily used to configure parameters to explore new system signals [22]. In this section, SDR is used to demodulate the PPP-B2b signal. Fig. 3 shows the block diagram of the receiver, which includes the acquisition module, tracking module, bitstream demodulation module and B2b-PPP information extraction module. The whole method is described as follows.

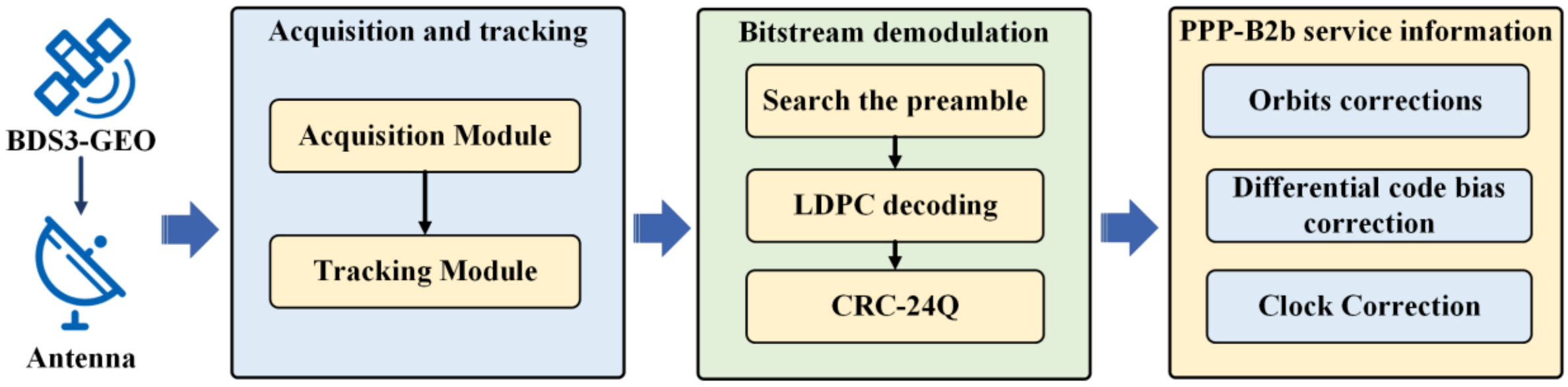

Fig. 3. Block diagram of the SDR of the PPP-B2b signal

### *A. ACQUISITION AND TRACKING MODULE*

After the raw data is successfully collected by SDR, the acquisition is carried out. The purpose of acquisition in this work is to determine visible satellites of GEO and calculate the coarse values of carrier frequency and code phase of the PPP-B2b signals [23]. To this end, the collected raw data are correlated with the local generated ranging code. In principle, there are two methods on the correlation, i.e. serial search acquisition in the time domain and the parallel search acquisition in the frequency domain is applied, in which Fourier transform is utilized to perform a transformation from the time domain into the frequency domain and the Doppler frequency shift and code phase are searched in the frequency domain in parallel [20], [22], [23].

It is worth mentioning that PPP-B2b signals are transmitted by the BDS-3 GEO satellites, which are about 35786 km away from the earth's surface, and located at 80°E, 110.5°E, and 140°E respectively [21]. Thus, the Doppler frequency shift caused by GEO satellite movement is kept in a small range, and the maximum Doppler frequency shift can be about 1.46 Hz every 1 Km/h even if the receiver is in the moving state [24]. Therefore, the search range of the Doppler frequency shift in the acquisition module of PPP-B2b is set within ± 1000 Hz, which accelerates the whole acquisition time.

The coarse value of the frequency Doppler shift and code phase is then by the acquisition module is put into the tracking module. The classic delay locked loop and phase locked loop is further used to precisely track the phase of carrier frequency and the ranging code. After the convergence of the tracking loops, the phase errors and code errors are kept within a small range [20], [22], [25]. The data demodulation is then carried out.

### *B. BITSTREAM DEMODULATION MODULE*

In order to decode PPP service information correctly, three steps are carried out, i.e., preamble searching 64-ary LDPC decoding and CRC. The methods are presented as follows:

#### 1) PREAMBLE SEARCHING

The purpose of preamble searching is to find the start of the PPP-B2b message. In PPP-B2b, preamble is the synchronization head designed as a fixed 16 bits unique word with the value of 0xEB90 in hexadecimal (i.e., 1110 1011 1001 0000 in binary). The preamble. It should be noted that, the preamble may occur in an inverse code in binary, i.e., in the value of 0001 0100 0110 1111, which is due to the 180° phase shift in the code tracking loop. Naturally, these two patterns can occur anywhere in the received data. So an additional check to authenticate the preamble must be carried out. Based on the design of every frame in the PPP-B2b signal, the repeated check on the occurrence of preamble in the bitstream from the tracking results is 1s (1000ms).

### 2) LDPC ENCODING AND DECODING

When the beginning of each frame is correctly detected with the assistance of the preamble information, the last 972 symbols of each frame is the message sequence of PPP-B2b. After decoded by LDPC (162,81), the PPP service information with a length of 486 bits can be extracted. The LDPC (162,81) encoding and decoding processes are introduced as follows.

Each frame of the PPP-B2b navigation message before error correction encoding has a length of 486 bits and the encoding scheme adopts 64-ary LDPC (162,81). Each codeword symbol is composed of 6 bits and defined in $GF(2^6)$ domain with a primitive polynomial of $p(x) = 1 + x + x^6$. A vector representation (MSB first) is used to describe the mapping relationship between non-binary symbols and binary bits **[19]**. Thus 486 bits will be converted into 81 codewords. The check matrix is a sparse matrix $\mathbf{H_{81,162}}$ of 81 rows and 162 columns defined in $GF(2^6)$ domain with a primitive polynomial of $p(x) = 1 + x + x^6$. It can be generated by the index matrix $\mathbf{H}_{\mathbf{81,162},\boldsymbol{index}}$ and element matrix $\mathbf{H}_{\mathbf{81,162},\boldsymbol{element}}$. The subsequent encoding steps are shown in Fig. 4.

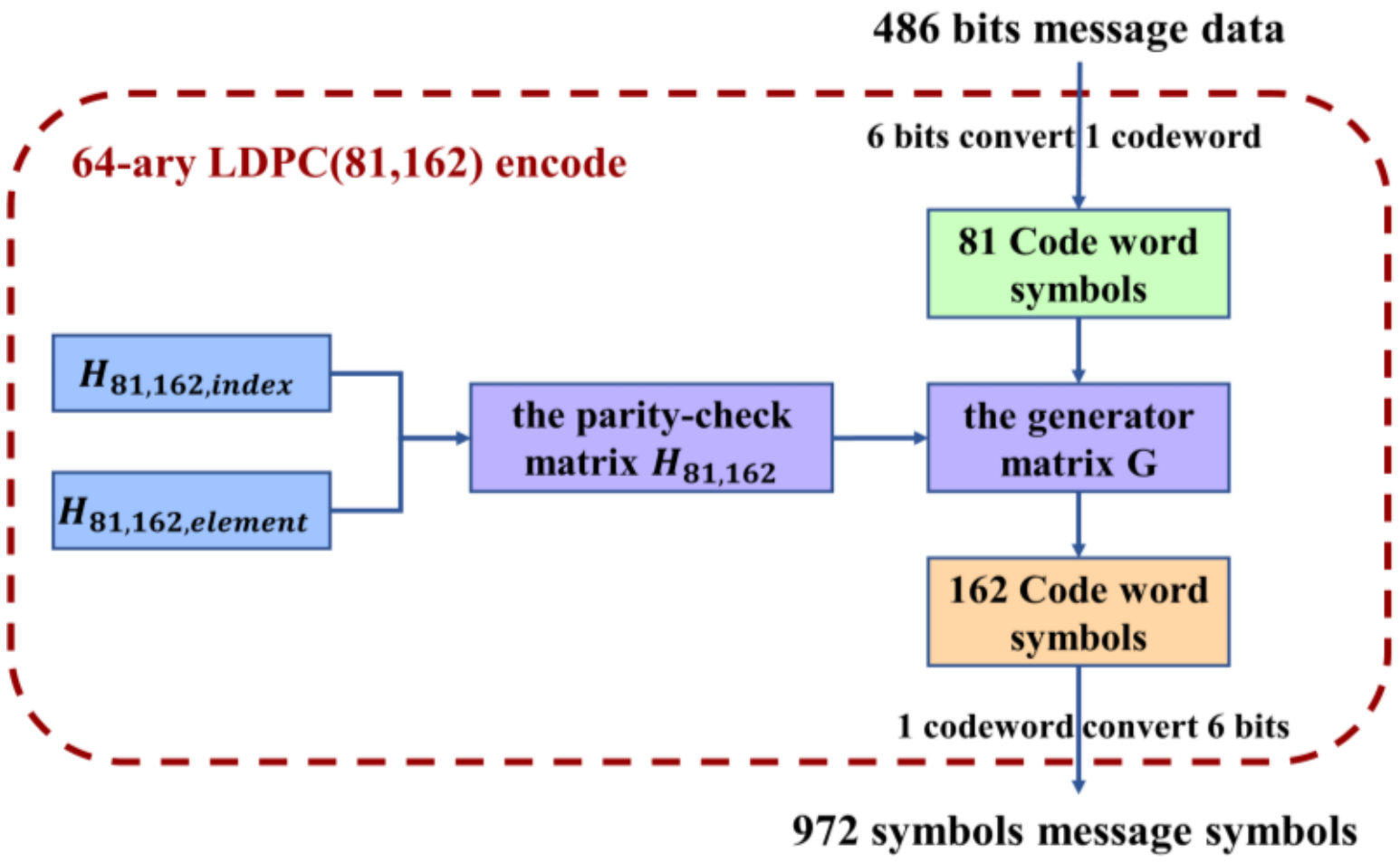

Fig. 4. 64-ary LDPC(81,162) encode process

As shown in Fig. 4, the generator matrix $\mathbf{G}$ can be calculated by the sparse matrix $\mathbf{H_{81,162}}$. By using generator matrix $\mathbf{G}$, the input sequence $m$ with a length of 81 can be encoded to obtain the codeword sequence with a length of 162. Finally, each codeword can be reconstructed back to 6 bits and the result will be converted into the navigation symbol sequence with a total length of 972 bits [21].

The codeword sequence encoded by LDPC (162,81) are denoted as:

$$\mathbf{c} = (\mathbf{c_0}, \mathbf{c_1}, \dots, \mathbf{c_{n-1}}), 0 \le n < 162 \tag{4}$$

After the modulated codeword sequence is transmitted through the signal channel, the receiver can get the corresponding sequence y as follow:

$$\mathbf{y} = (\mathbf{y_0}, \mathbf{y_1}, \dots, \mathbf{y_{n-1}}) \tag{5}$$

where $\mathbf{y_j} = (y_{j,0}, y_{j,1}, \dots, y_{j,r-1})$ is the received information corresponding to the $j^{th}$ codeword $c_j(c_j \in GF(2^r),\ r = 6, 0 \le j < n)$. The parity-check matrix H of the 64-ary LDPC(162,81) code can be used to check the correctness of the received sequence $\mathbf{y}$. the specific method is described as follows: A hard decision codeword $\hat{c} = (\hat{c}_0, \hat{c}_1, \dots, \hat{c}_{n-1})$ is obtained by making a hard decision on the received sequence y bit by bit. The check sum is calculated as $\mathbf{s} = \hat{\mathbf{c}}\mathbf{H}^\mathbf{T}$. If $\mathbf{s} = \mathbf{0}$, $\hat{\mathbf{c}}$ is outputs as the correct decoding result, otherwise $\hat{\mathbf{c}}$ is erroneous [21].

As shown in Figure 4, the generator matrix $\mathbf{G}$ can be calculated by the sparse matrix $\mathbf{H_{81,162}}$. By using generator matrix $\mathbf{G}$, the input sequence m with a length of 81 can be encoded to obtain the codeword sequence with a length of 162. Finally, each codeword can be reconstructed back to 6 bits and the result will be converted into the navigation symbol sequence with a total length of 972 bits [21].

*A. 64-ary LDPC(162,81) decoding*

The codeword sequence encoded by LDPC (162,81) are denoted as:

$$\mathbf{c} = (\mathbf{c_0}, \mathbf{c_1}, \dots, \mathbf{c_{n-1}}) ,0 \le \mathrm{n} < 162 \tag{4}$$

After the modulated codeword sequence is transmitted through the signal channel, the receiver can get the corresponding sequence $\mathbf{y}$ as follow:

$$\mathbf{y} = (\mathbf{y_0}, \mathbf{y_1}, \dots, \mathbf{y_{n-1}}) \tag{5}$$

where $\mathbf{y_j} = (y_{j,0}, y_{j,1}, \dots, y_{j,r-1})$ is the received information corresponding to the $j^{th}$ codeword $\mathrm{c_j}(\mathrm{c_j} \in \mathrm{GF}(2^r),\ r = 6, 0 \le j < n)$. The parity-check matrix H of the 64-ary LDPC (162,81) code can be used to check the correctness of the received sequence $\mathbf{y}$. the specific method is described as follows: A hard decision codeword $\hat{c} = (\hat{c}_0, \hat{c}_1, \dots, \hat{c}_{n\text{-}1})$ is obtained by making a hard decision on the received sequence y bit by bit. The check sum is calculated as $\mathbf{s} = \hat{\mathbf{c}}\mathbf{H}^{\mathbf{T}}$. If $\mathbf{s} = \mathbf{0}$, $\hat{\mathbf{c}}$ is outputs as the correct decoding result, otherwise $\hat{\mathbf{c}}$ is erroneous **[19]**.

The parity-check matrix $\mathbf{H}$ denotes the connection relationship between the check node (CN) and the variable node (VN), i.e., the reliability information can be transmitted between the connected CN and VN. The implementation of the reliability transmitting decoding algorithm can correct the received sequence $\mathbf{y}$ to estimated the transmitted codeword $\hat{\mathbf{c}}$. BeiDou official documents provide two iterative reliability transmitting decoding algorithms: (1) Extended Min-Sum Method; (2) Fixed Path Decoding Method. In this module, the Extended Min-Sum Method is used to estimate the transmission codeword $\hat{\mathbf{c}}$. The extended minimum sum decoding algorithm is described as follows [21]:

**Decoding algorithm**: The extended Min-Sum decoding algorithm

**Input**: 972 symbols PPP data sequence, the parity-check matrix H
**Output**: 486 bits PPP data sequence
**Initialize**: set the maximum number of iterations as $\mathrm{itr_{max}}$, itr = 0.

```
do
    calculate reliability vector L_j by 972 symbols PPP data sequence
    repeat each variable node VN_j then
        calculate the decision symbols ĉ_j and the reliability vector V2C
    until meeting the variable node updating rules
    calculate the check sum s = ĉH^T
end do
if s = 0 then
    return the decision symbols ĉ_j as output
end if
itr = itr+1
repeat
    if itr = itr_max then
        declare decoding failed
    end if
    repeat each check node CN_i then
        calculate the reliability vector C2V
    until meeting the check node updating rule
    repeat each variable node VN_j then
```

```
            calculate the decision symbols  ĉj  and the reliability vector   V2C
        until meeting the variable node updating rules
        calculate the check sum   s = ĉH^T
        itr = itr+1
    until   s = 0    return the decision symbols  ĉj  as output
```

The updating rule of variable nodes and check nodes, and the specific calculation method involved in the above algorithms can refer to the PPP-B2b signal official document [21].

#### 3) CRC

After the 64-ary LDPC (162,81) decoding, the CRC will be processed. The CRC check is performed by bit for the 6 bits message type and the 456 bits message data. The lowest 24 bits are CRC code, as shown in the upper layer of Fig. 2.

The implementation of CRC is CRC-24Q and its generator polynomial is [26]:

$$g(x) = \sum_{i=0}^{24} g_i x^i \tag{6}$$

where, $g_i = \begin{cases} 1, i = 0,1,3,4,5,6,7,10,11,14,17,18,23,24 \\ 0, else \end{cases}$

Furthermore, $g(x)$ can be expressed as follows [26]:

$$g(x) = (1 + x)p(x) \tag{7}$$

where, $p(x) = x^{23} + x^{17} + x^{13} + x^{12} + x^{11} + x^9 + x^8 + x^7 + x^5 + x^3 + 1$.

By dividing decoded sequences with the generator polynomial $g(x)$, the residue should be all "0". However, due to the channel effect on signal transmission, there might be errors in the decoded sequence. Therefore, the results of CRC can be used to determine whether the decode sequence is correct or not.

### C. PPP INFORMATION EXTRACTION MODULE

After LDPC decoding and CRC, the PPP service information can be extracted. The PPP service information mainly includes information content such as satellite orbit corrections, differential code bias, and satellite clock corrections. As stated in Section II-B, different information content is indicated by different message types. Before the PPP service information can be effectively applied to the receiver, different information content has to be matched first based on the information of a group of Issue of Data (IOD). For details of the IODs, please refer to BeiDou official documents [21].

## IV. EXPERIMENT AND RESULTS

To verify the feasibility of the PPP-B2b SDR designed in Section III, and to explore the characteristics of the PPP-B2b signal, the field tests were carried out on the roof of a building on the campus of Wuhan University on September 8, 2020.

### A. FIELD EXPERIMENTS

The testing scenario and test bench are shown in Fig. 5. As shown in Fig. 5(a), an open environment is selected for the signal collection to avoid the multipath impact. In Fig. 5(b), the raw data is recorded by LabSat 3 wideband, which is a radio frequency (RF) signal recording and playback system. In LabSat 3 wideband, the supported frequency varies from 1.2 GHz to 1.6 GHz, the sampling rate from 10 MHz to 58 MHz and the quantization precision ranges from 1, 2 to 3-bit in I and Q digital signal. In this field test, the parameters of receiver are configured in Table III.

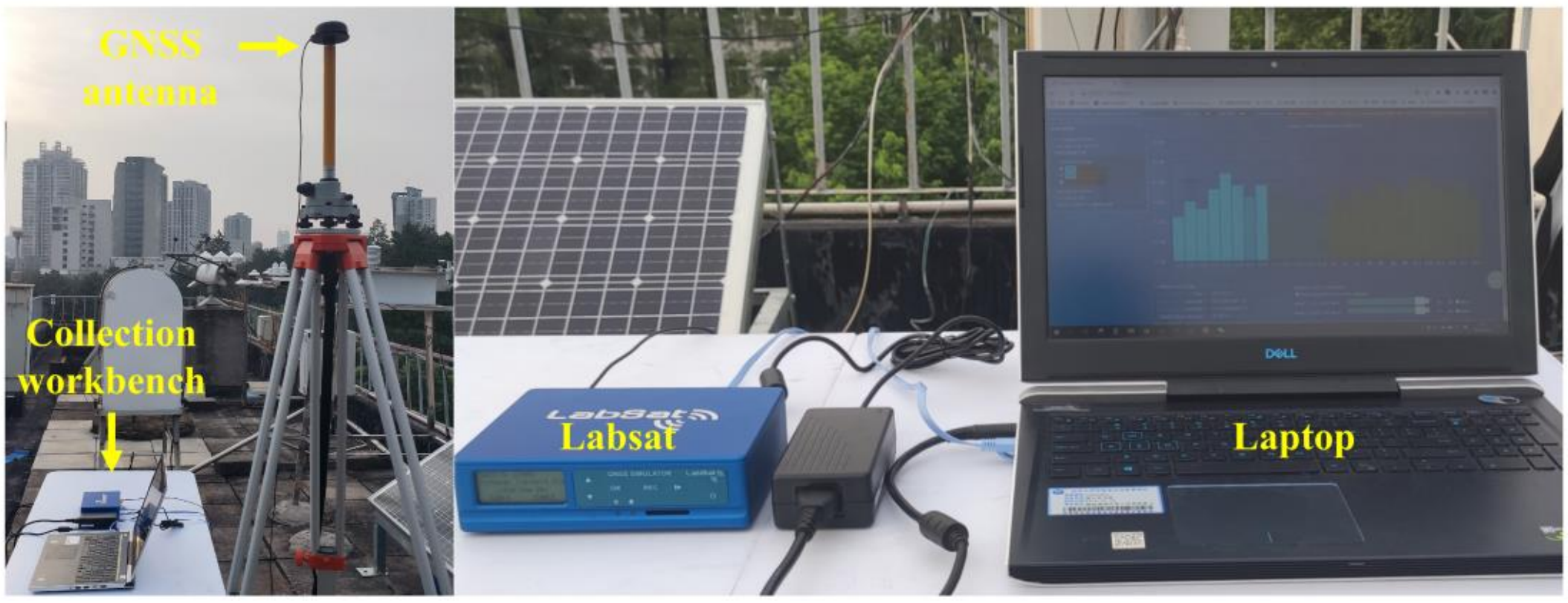

(a) (b)

Fig. 5. The testing scenario and test bench of the experiment. (a) The GNSS antenna and collection workbench were arranged on the roof of the building of Wuhan University. (b) The detail of the collection workbench. The workbench is mainly composed of two parts: Labsat 3 wideband and laptop. The laptop is used to configure the parameters of Labsat 3 wideband.

TABLE III

THE PARAMETERS OF CONFIGURATION DURING THE DATA COLLECTION.

| **Center frequency (MHz)** | **Number of channels** | **Bandwidth (MHz)** | **Sampling rate (MHz)** | **Quantization (bit)** | **Time (s)** |
|---|---|---|---|---|---|
| 1575.42<br>1207.14 | 2 | 30 | 30.69 | 2 | 3900 |

### B. RESULTS OF ACQUISITION AND TRACKING

In the test, the satellites with PRNs of 19, 21, 22, 34, 35, 39, 40, and 44 were acquired using the B1C signal, the satellites with PRNs of 59, 60, and 61 are acquired by B2b signal and the satellites with PRNs of 2, 5, 6, 12, 13, 15, and 29 by the L1 C/A signal. The code phase and Doppler frequency shift of the acquired satellites of BDS3 and GPS are presented in Table IV. As is clear, in general, the Doppler frequency shift of PPP-B2b, which is from the GEO, is much smaller than that from the BeiDou B1C and GPS L1 C/A, which are either from MEO or IGSO.

TABLE IV

THE CODE PHASE AND DOPPLER FREQUENCY SHIFT OF THE ACQUIRED SATELLITE.

| **DBS3-PRN** | **BIC（19-44）** | | | | | | | | **PPP-B2b（59-61）** | | |
|---|---|---|---|---|---|---|---|---|---|---|---|
| | 19 | 21 | 22 | 34 | 35 | 38 | 39 | 44 | 59 | 60 | 61 |
| Doppler (Hz) | 2465 | -2001 | 479 | -3157 | 2016 | -504 | -1352 | -1084 | -29 | 39 | -71 |
| Code Offset (Sample) | 82085 | 68005 | 5702 | 37910 | 33463 | 82864 | 17990 | 7634 | 4283 | 6819 | 7982 |
| GPS-PRN | 2 | 5 | 6 | 12 | 13 | 15 | 29 | | | | |
| Doppler (Hz) | 10 | 595 | 2088 | 2478 | -2624 | -3737 | -2566 | | | | |
| Code Offset (Sample) | 1888 | 129 | 7763 | 8962 | 1245 | 9751 | 529 | | | | |

Fig. 6 shows the tracking results of PPP-B2b signals from PRN59.From Fig. 6(a), it is observed that, the signal power is mainly concentrated in the I phase, while the Q phase only contains the noise, which is consistent with the theoretical design [20]. Fig. 6(b) shows the bitstream from I phase and the data transmission rate of the bitstream is 1000 sps.

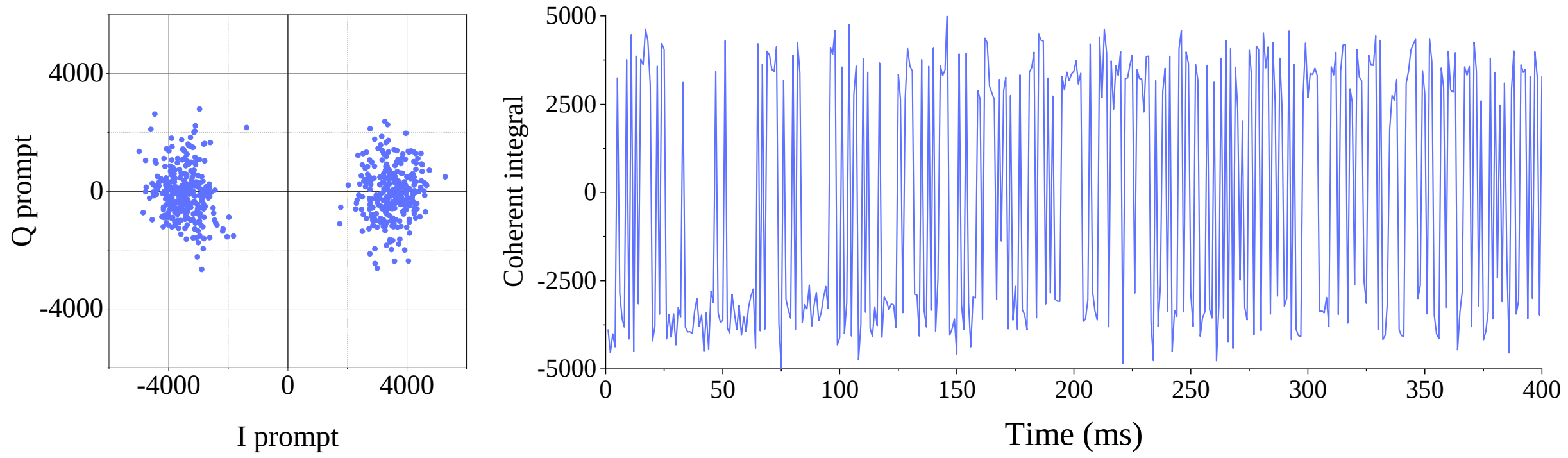

(a) (b)

Fig. 6. Tracking results of PRN59 (PPP-B2b). (a) The power distribution of I and Q phase of tracking results. (b) The bitstream from I phase. The transmission rate of the bitstream is 1000 sps.

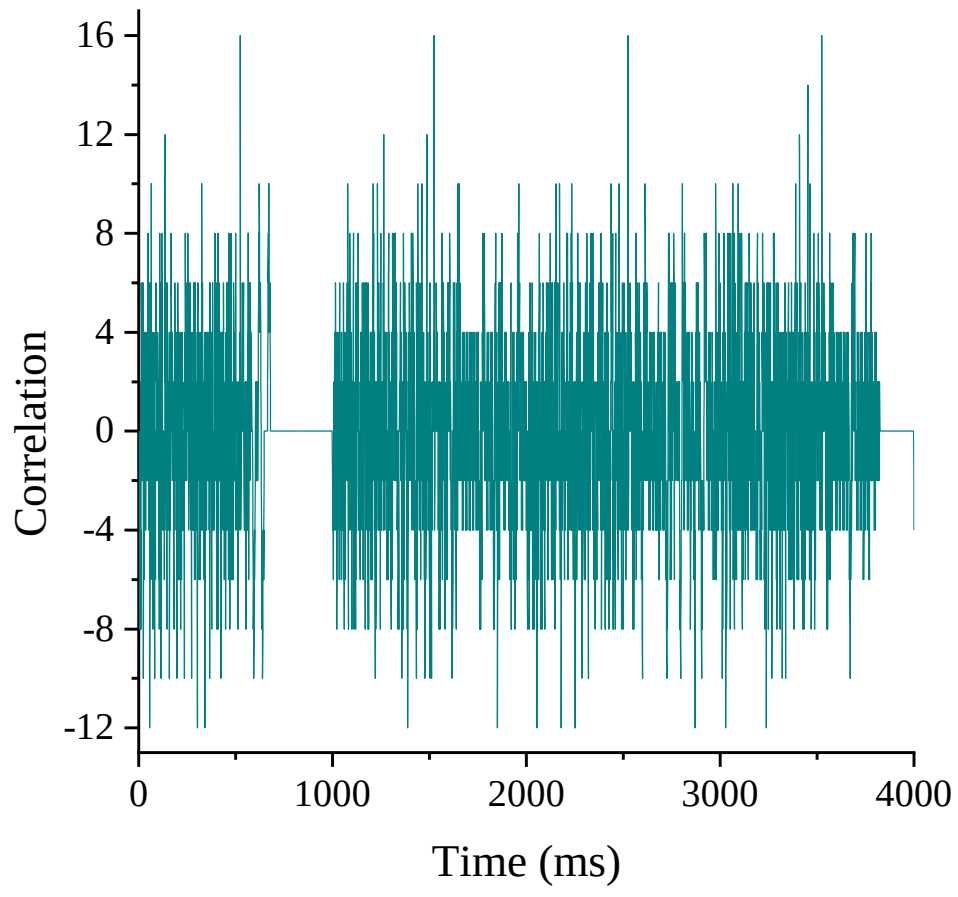

Fig. 7. Correlation between 4000ms message sequence and 16-bit preamble (the peak value indicates the start position of the frame)

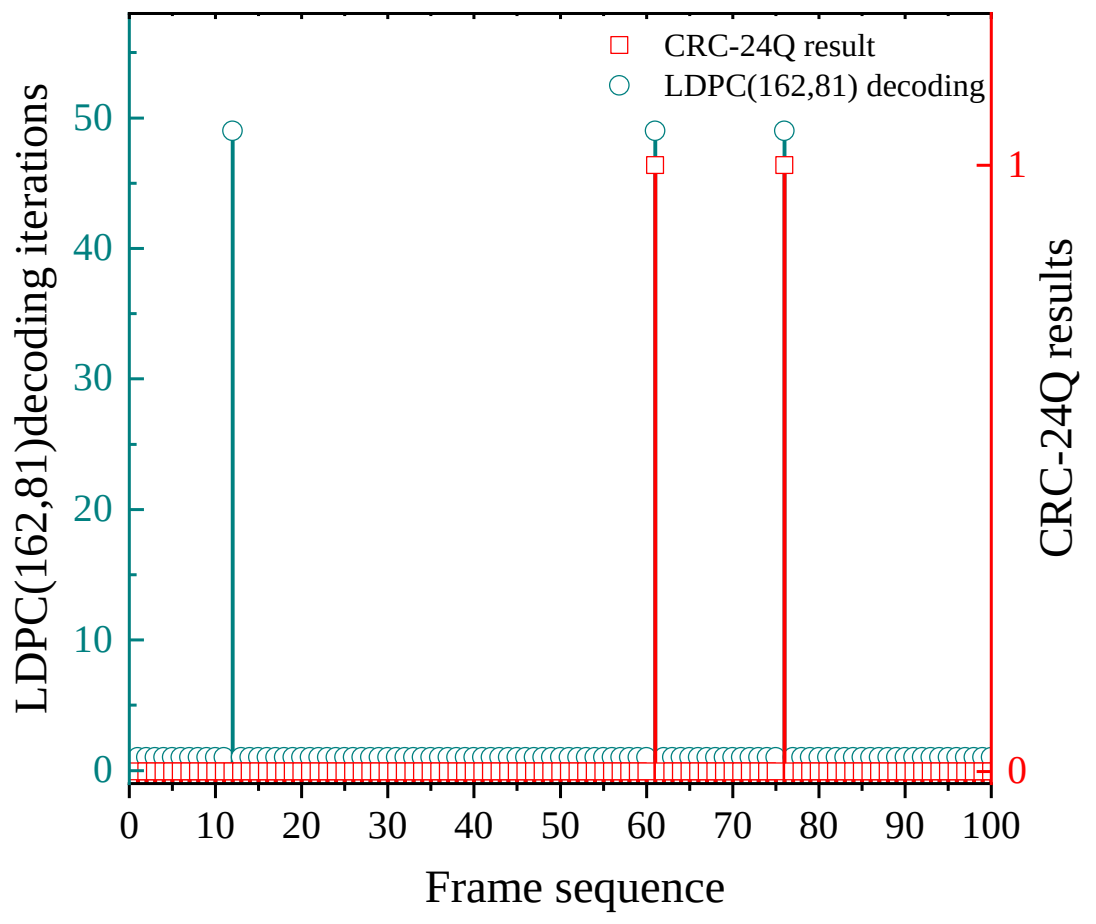

Fig. 8. The results of 64-ary LDPC(162,81) decoding and CRC of 100 frame data

### *C. Bitstream Demodulation Results*

Fig. 7 shows the correlation results of the preamble search. We can find that there are 4 synchronous heads with the correlation peak value of 16, and by calculation, their intervals are all 1000 ms, which suggests the successful detection of the start of the frame. Fig. 8 shows the results of the LDPC decoding and CRC from 100 frame sequence, in which, the number of the frame sequence vs. the iteration times of the LDPC decoding is plotted.

As shown in Fig. 8, in the 100 frames of data, most of the frames passed the CRC after the LDPC decoding with 1 iteration. In only three cases, the number of iterations reaches the maximum number of iterations, two of which fail to pass the CRC due to the erroneous decoded data. The correctly decoded frame data sequences are then processed to extract PPP service information. However, it is unfortunate that, the current version of the PPP-B2b space interface document has not described the design of a complete set of PPP information frame structures. In this work, we summarize the frame layout format from measurement data in the field tests in the following sections.

### *D. ARRANGEMENT OF FRAME TYPES*

The length of each frame message is 486 bits, of which the first 6 bits are the message type of the frame. However, the BeiDou official document does not mention the repetition period of each frame and the frame arrangement of a set of PPP service messages, so we output the highest 6-bit frame message type, and the results are shown in Fig. 9.

Fig. 9 shows the frame arrange of the epoch moment from 17998 to 18046. After message type 1 has been broadcast, message type 4 and message type 3 are broadcast alternately, and information type 4 is broadcast continuously. When message type 3 has been broadcast for one round, message type 2 is added. When the message type 2 has been broadcast for one round, the message type 63 is broadcast in the remaining time. By summarizing the results of many times, we have defined a set of PPP-B2b frame format as follows:

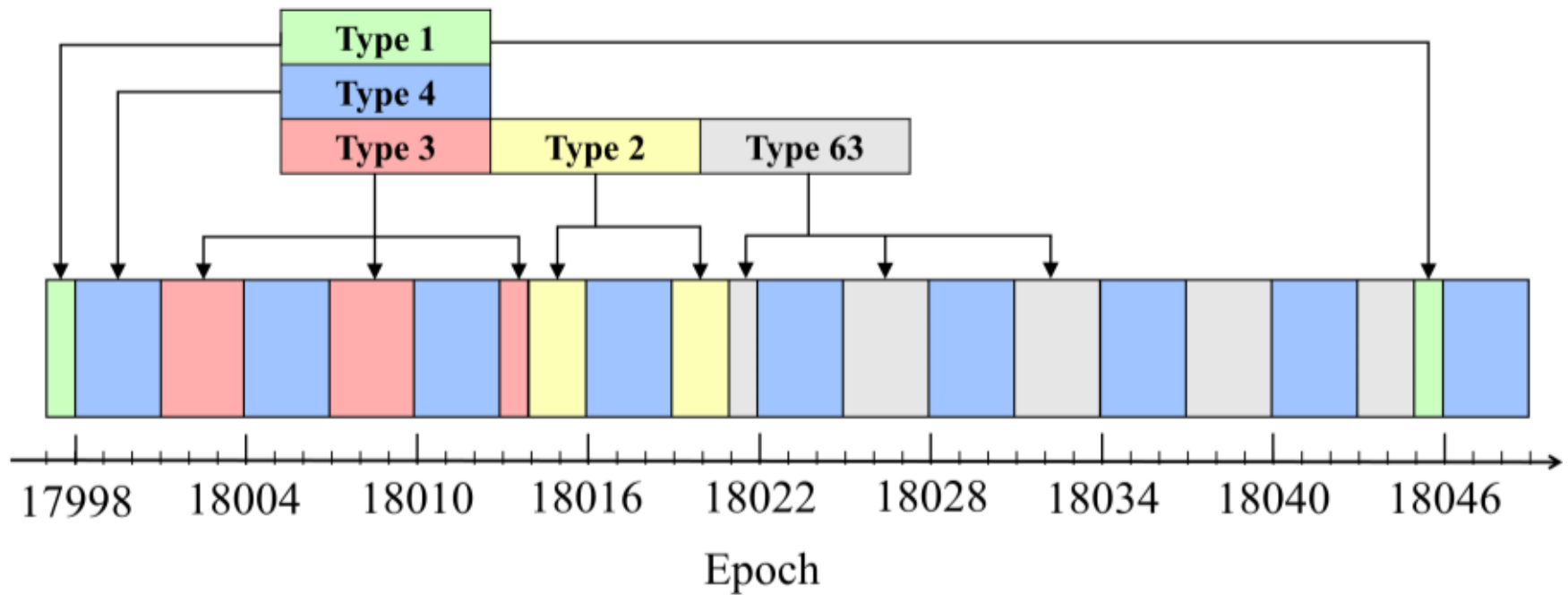

Fig. 9. Frame arrangement structure

1) Taking the time of message type 1 as the starting position of a complete PPP service information, the broadcast period is 48 s;
2) The remaining 47 s broadcast message types are 2, 3, 4, 63 (and the defined message types 5, 6 and 7 have not yet appeared in the measured data);
3) The duration of the remaining 47 s broadcast message type can be regarded as 3 s as an output unit, there are 15 times, and the 16th time is 2 s;
4) Message type 4 is regarded as one output type, and another output type is composed of message types 2, 3 and 63. The two output types are alternately broadcast in one output unit;
5) Message type 4 is continuously broadcast, and the epoch is constantly updated;
6) Message types 2 and 3 are only broadcast once in a set of PPP information, and the corresponding output unit broadcast message type 63 at the remaining time fills in the blank position, and the message type 63 is broadcast in the last 2 s.

In addition, when extracting PPP information, we found that the update time and corresponding epoch of corrections of PPP-B2b signals are different. Their first epoch moment is the same as the epoch moment of message type 1, and the update rate of the epoch moment of corresponding message type 4 is also 6 s. Message types 2 and 3 use the same epoch moment but are 7 s slower than the epoch moment of message type 1, and their update rate is 48 s. The decoding results of PPP service information will be shown in the next.

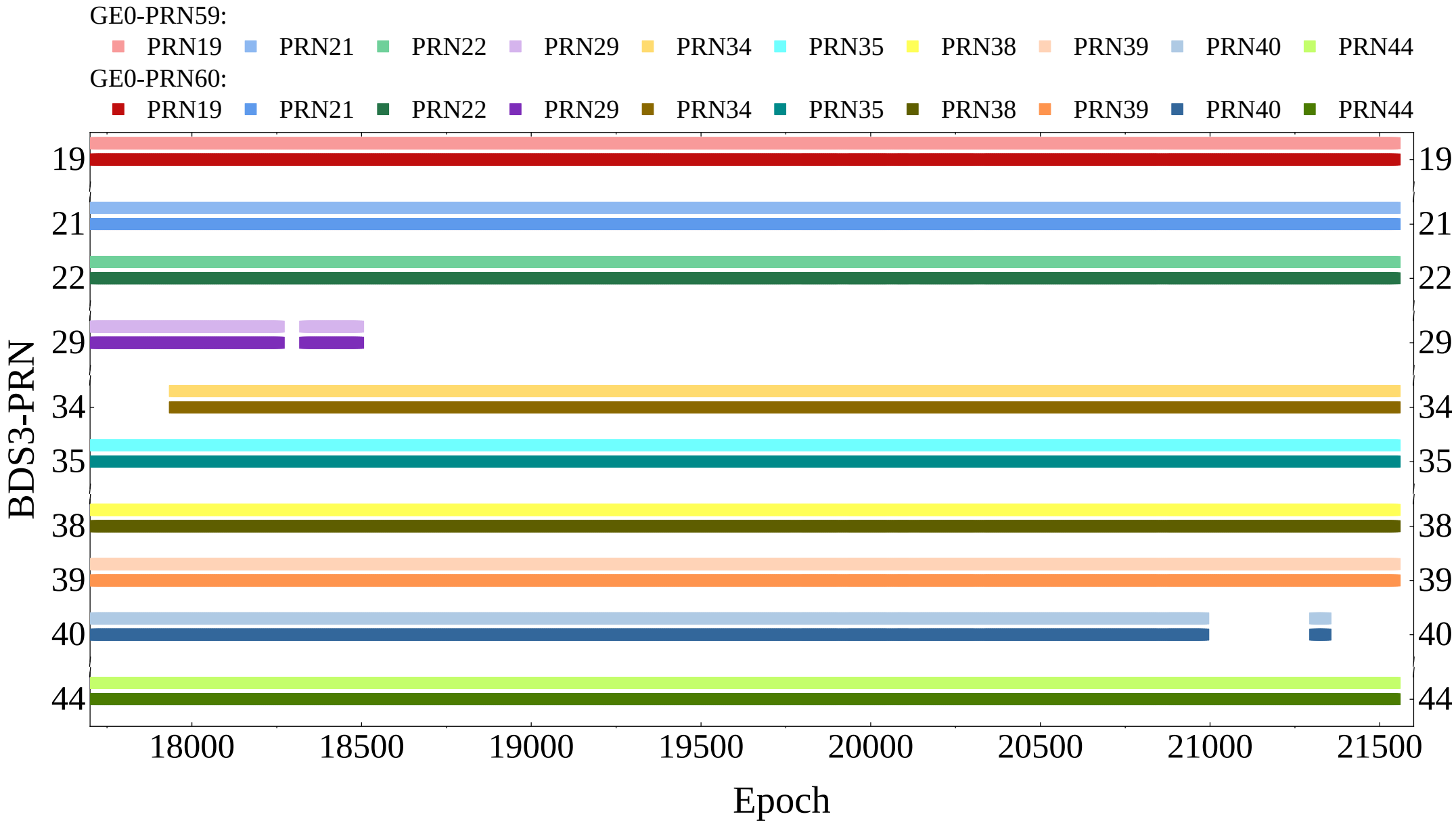

Fig. 10. The PRN and duration of BDS3 satellites with valid corrections

### *E. Decoding Results of PPP Service Information*

To obtain the PPP service information, the digital signal data were processed according to the proposed method in this experiment. It was found that the satellites mask provided by the message type 1 in the PPP-B2b service information includes all satellites of BDS3 and GPS, that is, the satellites of BDS3 and GPS corresponding masks are all set to "1". However, due to the corrections of some satellites are larger than the effective value set by the document, their corrections are unavailable. The duration of PPP service information providing effective correction for BeiDou satellites (BDS3) is shown in Fig. 10. The PPP service information is provided by the GEO satellites (PRN-59 and PRN-60).

From Fig. 10 we found that the PRN of satellites with effective corrections are consistent with the PRN of satellites in the Asia-Pacific region at the moment, that is, they are similar to the PRN of satellites in the results of the acquisition module. Fig. 11 shows the duration of the PPP service information served for the satellites of GPS.

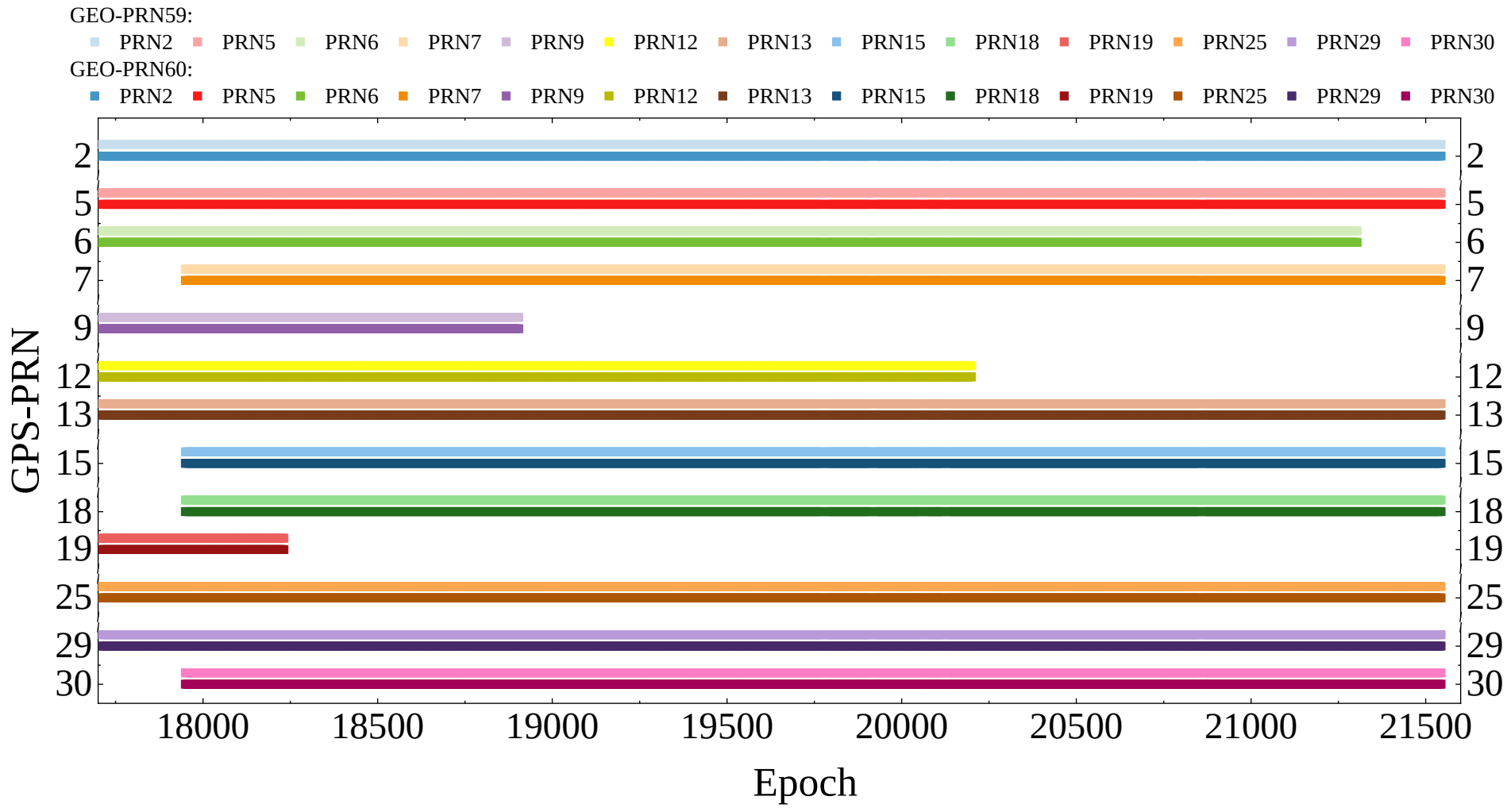

Fig. 11. The PRN and duration of GPS satellites with valid corrections

As shown in Fig. 11, the PPP service provided for GPS is similar to the PPP service provided for BDS. In addition, the PPP service information provided by PRN-59 satellite is the same as that provided by PRN-60 satellite. Next, these effective corrections are analyzed. Fig. 12 shows the effective orbit corrections of BDS-B1C satellites, with a time range of about 65 mins.

As shown in Fig. 12, we found that the orbit corrections of BDS satellites are in the range of centimeter to decimeter. From the epoch moment of view, when the BDS3-B1C navigation message is updated, that is, epoch moment equal to 18000, the PPP-B2b service information will be updated immediately. At that moment, most of the satellite orbit corrections (radial, along and cross) have different degrees of step-change rather than continuity. In a complete epoch moment range, that is, the epoch moment is from 18000 to 21549, the changes of orbit corrections are relatively continuous. From the satellite type, there is no significant difference between the corrections of MEO and IGSO. But we found that the standard deviation of the radial direction of MEO satellites are far less than that of the along and cross direction, and the standard deviation of IGSO satellites are similar, but the differences are not as significant as that of MEO satellites. Fig. 13 shows the effective orbit corrections of GPS satellites, with a time range of about 65 mins.

From the comparison of Fig. 13 and Fig. 12, it can be found that the change of orbit corrections of GPS satellites is different from that of BDS, and the changes are smooth and continuous without step changes when the navigation message is updated. The orbit corrections of GPS satellites are much greater than the BDS correction. The orbit corrections of GPS satellites are in the range of decimeter to meter, which is much larger than that of BDS satellites.

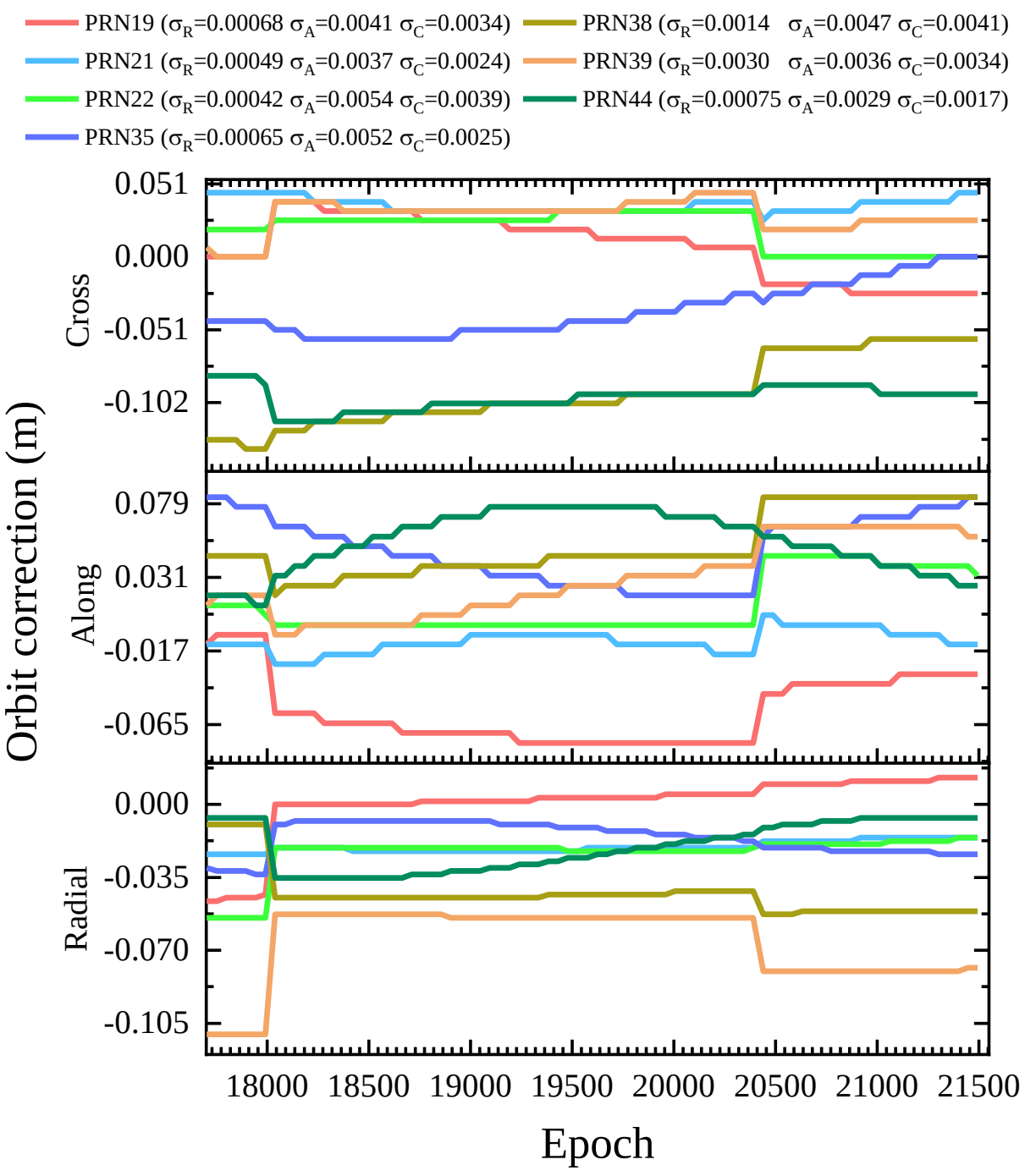

Fig. 12. Changes in orbit corrections (radial, along, cross) of BDS3 BIC. There are seven BDS3 satellites with orbit corrections, and the epoch is from 17710 to 21549.Top: cross orbit correction, middle: along orbit correction, bottom: radial orbit correction.

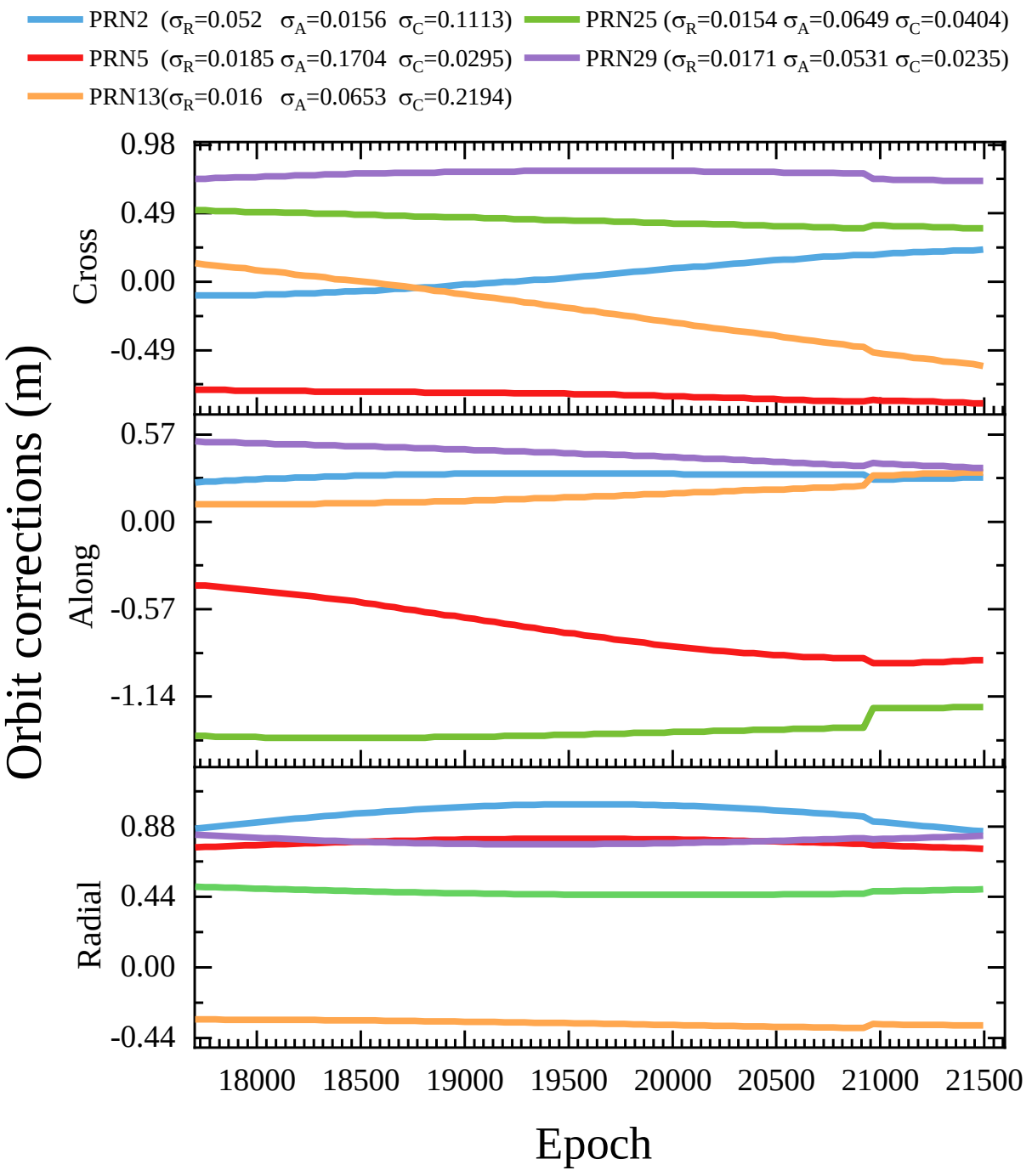

Fig. 13. Changes in orbit corrections (radial, along, cross) of GPS L1. There are five GPS satellites with orbit corrections, and the epoch is from 17710 to 21549. Top: cross orbit correction, middle: along orbit correction, bottom: radial orbit correction.

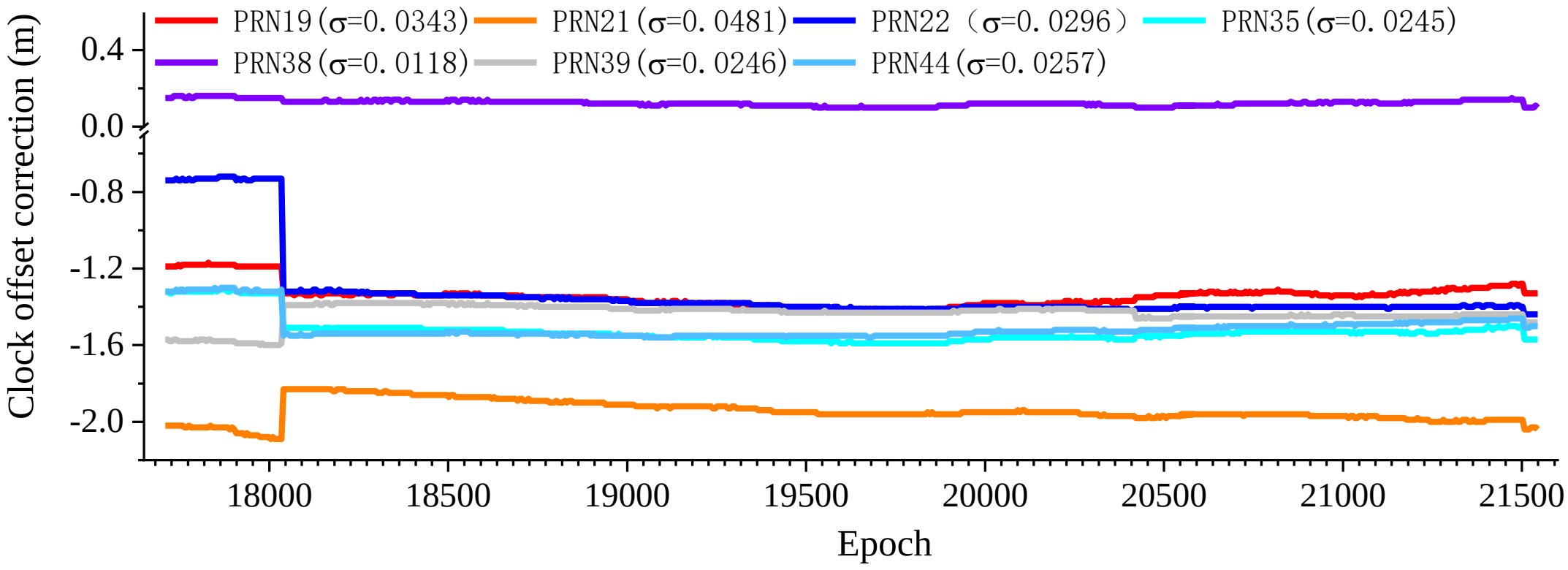

Fig. 14. Changes in clock corrections of BDS3 satellites. There are seven BDS3 satellites with clock offset corrections, and the epoch is from 17710 to 21549. The PRN of satellite in Fig. 14. correspond to those in the Fig. 12.

In Fig. 14, the clock corrections and their standard deviation of each BDS satellite over a complete epoch range are shown. From Fig. 14, we can know that the clock correction of almost all BDS satellites is within the meter range. For the change of clock corrections, when the BDS3-B1C navigation message is updated, the clock correction of most satellites also has a step change to varying degrees. But in a complete epoch, the clock correction changes of each satellite are very small reflected from their standard deviation, and their changes are inconsistent.

Fig. 15 shows the clock correction of GPS satellites over the entire epoch. The clock corrections are in the range of decimeter to meter level. For the changes in the clock correction of satellites, three obvious situations are different from those of BDS satellites in Fig. 15. It can be seen that when the navigation message changes, there is no step-change in clock correction. The changes in the clock corrections of the satellites are consistent during the entire period, which is reflected from their standard deviation. In addition, we found that the corrections of PRN-5 are all “0” in some periods. The reason for this situation may be that the clock correction is too close to "0", resulting in an insufficient number of bits to express it. The differential code bias provided by PPP-B2b is the pseudo-range code biases between the ranging signal and the clock offset reference signal adopted by the corresponding system. Users need to correct the differential code bias accordingly when they are using the signals other than the reference signal, otherwise, the convergence time of precise point positioning may be affected [21]. Refer to the PPP-B2b signal interface document to check the specific content of the signal component corresponding to the differential code bias and the signal-receiving mode of that component [21]. Among them, PPP-B2b service only provides the differential code bias of BDS3-B1C, the corresponding model including 0,1,2,4,5,7,8,12. During this epoch, the signal and tracking mode and corresponding corrections have no change.

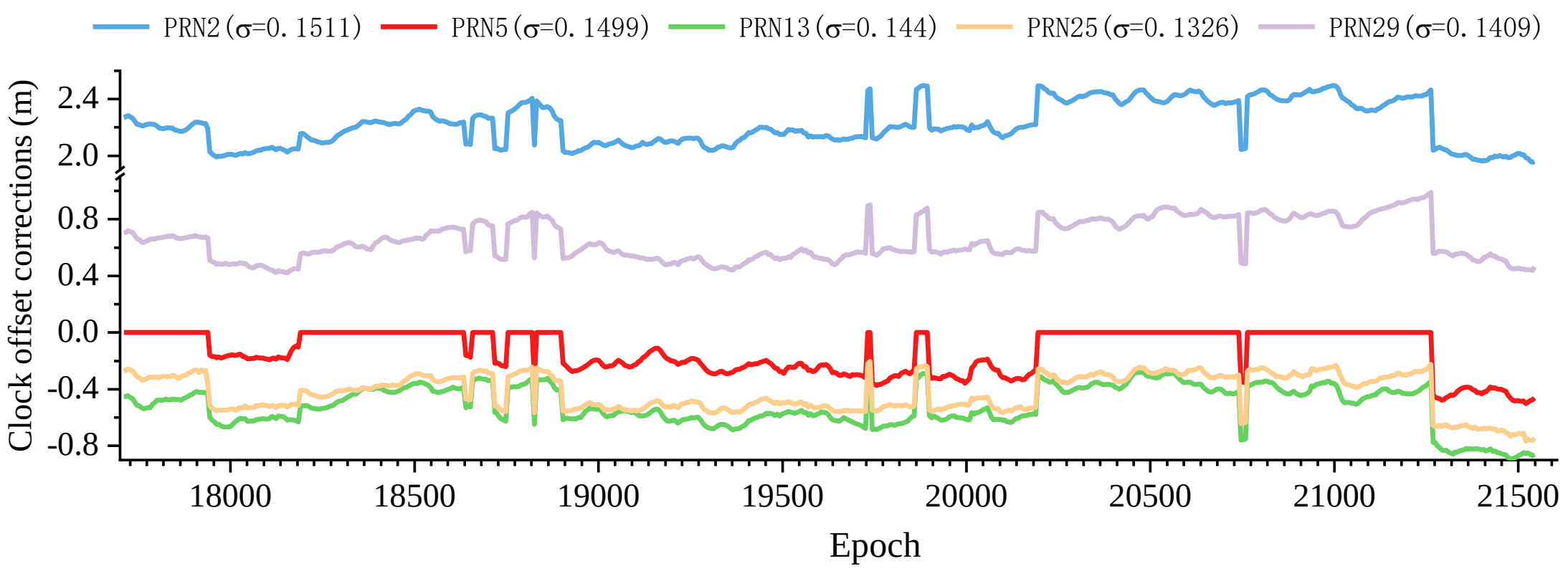

Fig. 15. Changes in clock corrections of GPS satellites. There are five GPS satellites with clock offset corrections, and the epoch is from 17710 to 21549. The PRN of satellite in Fig. 15. correspond to those in the Fig. 13.

### F. INTEGRITY OF PPP SERVICE INFORMATION

Finally, we found that the remaining satellites, which have not matched the complete PPP service, often have some cases of unavailable corrections or data duplication, as shown in Fig. 11 and Fig. 12. The situation that the epoch time broadcasted by the GEO satellite has not been updated due to its own clock problem. We classified the above-mentioned problems as abnormal problems of the GEO satellite PPP service. We analyzed the PPP service performance of two satellites GEO (PRN-59) during the collection period. The detailed results are shown in Table V.

From the results in Table V, we find that the completeness of the orbit correction of almost all satellites can reach 100%, and the completeness of the clock correction can reach more than 99%. Only PRN-29 and 40 of BDS3 have slightly worse completeness of clock correction and orbit correction than other satellites. The main reason for this situation is that the PPP service information of the two satellites is discontinuous in time, and the orbit correction exceeds the effective range. We treat it as a service exception. The PPP services of the two satellites stopped broadcasting when the epoch time was 18525 and 21357, respectively. Therefore, we believe that when PPP-B2b stops serving a satellite, its PPP service information will be unstable. Users who using RTPPP may need to prepare to stop the positioning results of the satellite as soon as possible to prevent the impact on the RTPPP results. In addition, the clock corrections of all satellites have anomalies of different durations. There are two abnormal situations. In most situations, the correction information is not updated, that is, the epoch of the frame is still the epoch of the previous frame.

The content of the next frame is broadcast as usual, and the interval between the two epochs is 12 s. Another situation is that the clock correction exceeds the effective range. However, clock correction has the characteristics of fast update speed and small changes. The user can choose to continue using the clock correction of the previous frame until a new clock correction is extracted. In general, the PPP information service provided by the PPP-B2b signal is very stable. At the same time, it has established PPP services for BDS3 and GPS, with the characteristics of the short broadcast cycle and fast update rate. It is also possible to use the PPP information of three GEO satellites to check each other without excessively relying on one satellite.

TABLE V

THE PARAMETERS OF CONFIGURATION DURING THE DATA COLLECTION.

| PRN | Epoch time (s) | Total time (s) | Abnormal time (s) and completeness ratio of orbit corrections | | Abnormal time (s) and completeness ratio of clock corrections | |
|---|---|---|---|---|---|---|
| **BDS3** | | | | | | |
| 19 | 17710 - 21549 | 3839 | 0 | 100% | 30 | 99.22% |
| 21 | 17710 - 21549 | 3839 | 0 | 100% | 30 | 99.22% |
| 22 | 17710 - 21549 | 3839 | 0 | 100% | 30 | 99.22% |
| 29 | 17710 - 18525 | 815 | 48 | 94.11% | 102 | 87.48% |
| 34 | 17950 - 21549 | 3599 | 0 | 100% | 30 | 99.17% |
| 35 | 17710 - 21549 | 3839 | 0 | 100% | 30 | 99.22% |
| 38 | 17710 - 21549 | 3839 | 0 | 100% | 30 | 99.22% |
| 39 | 17710 - 21549 | 3839 | 0 | 100% | 30 | 99.22% |
| 40 | 17710 - 21357 | 3647 | 288 | 92.1% | 366 | 89.96% |
| 44 | 17710 - 21549 | 3839 | 0 | 100% | 30 | 99.22% |
| **GPS** | | | | | | |
| **PRN** | **Epoch time (s)** | **Total time (s)** | **Abnormal time (s) and completeness ratio of orbit corrections** | | **Abnormal time (s) and completeness ratio of clock corrections** | |
| 2 | 17710-21549 | 3839 | 0 | 100% | 30 | 99.22% |
| 5 | 17710-21549 | 3839 | 0 | 100% | 30 | 99.22% |
| 6 | 17710-21309 | 3599 | 0 | 100% | 90 | 97.50% |
| 7 | 17950-21549 | 3599 | 0 | 100% | 30 | 99.17% |
| 9 | 17710-18909 | 1199 | 0 | 100% | 66 | 94.50% |
| 12 | 17710-20205 | 2495 | 0 | 100% | 84 | 96.63% |
| 13 | 17710-21549 | 3839 | 0 | 100% | 30 | 99.22% |
| 15 | 17950-21549 | 3599 | 0 | 100% | 30 | 99.17% |
| 18 | 17950-21549 | 3599 | 0 | 100% | 30 | 99.17% |
| 19 | 17710-18237 | 527 | 0 | 100% | 48 | 90.89% |
| 25 | 17710-21549 | 3839 | 0 | 100% | 30 | 99.22% |
| 29 | 17710-21549 | 3839 | 0 | 100% | 30 | 99.22% |
| 30 | 17950-21549 | 3599 | 0 | 100% | 30 | 99.17% |

## V. DISCUSSIONS

In this section, we will discuss the steps that can be optimized in the decoding process, predict and delete sequences with decoding errors, to reduce the occupation of computing resources and improve the decoding efficiency. Finally, for the PPP-B2b service information that has been extracted many times, some of the content that is not clearly defined in the official documents is summarized. One part has been explained in Section 4, and the other parts will be discussed in this section.

### A. OPTIMIZATION OF THE DECODING PROCESS

As mentioned in Section III, after finding the starting position corresponding to the synchronization head, their interval is used as the other evidence for judging the synchronization head. From the perspective of bit information arrangement, the last 6 after the preamble can be converted into satellite number (PRN), as shown in Fig. 2. It corresponds to the acquired ranging code number. It can be used as a secondary check to find the correctness of the synchronization head. Therefore, the method of finding the starting position of the frame includes a preliminary determination determining that the interval between two consecutive maximum correlation values is 1s as a preliminary determination. And a second determination that whether the satellite number (PRN) extracted by the symbol sequence should be consistent with the satellite number (PRN) obtained by the acquisition module. If the synchronization head is the inverse code, the satellite number extracted by the symbol sequence is also inverted accordingly, that is, the sum of it and the satellite number (PRN) obtained by the acquisition module should be 63.

In addition, we found that when the synchronization head detection is correct for the real signal data, the decoded result is still unusable. According to the above situation, by detecting the 6-bit satellite number (PRN), whether it matches the satellite number (PRN) obtained by the acquisition module, or matching after flipping. If both conditions do not match, it means that the corrections extracted by this sequence cannot be used. The problem may be caused by the tracking module. In order to reduce the occupation of computing resources, this sequence can be deleted.

### B. PROGRESS OF PPP-B2b

We have summarized some contents that were not clearly defined in the PPP-B2b official document by the BDS, including the progress of PPP-B2b service for the GNSS system and some other characteristic (as of September 8, 2020).

1) The PPP-B2b signal has provided real-time PPP services for the two GNSS systems, BDS3 and GPS.
2) Since the message type 1 provides satellites mask information, the message type 1 appears as a complete set of PPP service information broadcast time unit, with a total duration of 48 s, that is, the broadcast duration of a complete set of PPP service information is 48 s.
3) The PPP-B2b signal currently only provides PPP services to satellites acquired by BDS3-B1C and GPS signals covering the Asia-Pacific region. That is, which satellites (BDS3 and GPS) the user can acquire in the Asia-Pacific region can receive their corresponding PPP service.
4) The PPP service information is broadcast by the BDS3-GEO satellites, PRN is 59, 60, and 61 respectively. The three satellites jointly broadcast the same PPP service information.
5) When the collection conditions are limited, users can check and splice the PPP service information obtained by the three GEO satellites to restore the complete PPP service information.
6) The currently designed useful message types are 1-7. In actual situations, users can only receive message type 1-4.
7) Currently, the differential code bias corrections from message type 4 only provide services for BDS3 satellites.
8) The epochs of various PPP service corrections are 10-20 s slower than that of B1C navigation message.

## VI. CONCLUSIONS AND FUTURE WORKS

This paper aims at acquiring PPP services information from PPP-B2b signal. A PPP-B2b SDR is implemented, which includes the acquisition and tracking module, the bitstream decoding module, and the PPP service information extraction module. To verify the feasibility of SDR, field experiments are carried out, in which, the time-varying attributes of the corrections of BDS and GPS are evaluated, and the integrity and stability of the PPP information service were analyzed. The results also showed that the PPP-B2b signal can stably provide PPP services for satellites in the Asia-Pacific region, including centimeter-level to decimeter-level orbit corrections and meter-level clock corrections for BDS satellites. In the meanwhile, PPP services provide decimeter-level to meter-level orbit correction and meter-level clock correction for GPS satellites. Finally, detection tips for bitstream availability in SDR are proposed, and some content which is not defined in the official document, such as the PPP-B2b frame arrangement scheme, various correction update cycles and PPP-B2b's progress in providing PPP service are discussed. Related to the future works, the implementation of the decoded PPP-B2b service information directly on the SDR receiver will be considered and the positioning accuracy and integrity will be further compared in detail.

**ACKNOWLEDGE** This research was funded by the National Key Research and Development Program with Project No. 2018YFB0505400, and the Natural Science Fund of Hubei Province with Project No. 2018CFA007.